\documentclass[iop]{emulateapj}
\usepackage{amsmath}
\usepackage{natbib}

\catcode`_=\active
\newcommand_[1]{\ensuremath{\sb{\mathrm{#1}}}}
\newcommand{\icm}{cm$^{-1}$}

\newcommand{\be}[1]{\begin{equation} \label{eq:#1}}
\newcommand{\ee}{\end{equation}}
\newcommand{\ba}[1]{\begin{eqnarray} \label{eq:#1}}
\newcommand{\ea}{\end{eqnarray}}

\newcommand{\pref}{\protect\ref}

\newcommand{\solrad}{\ifmmode{R}_{\rm S}\else${R}_{\rm S}$\fi}
\newcommand{\solmas}{\ifmmode{M}_{\rm S}\else${M}_{\rm S}$\fi}

\newcommand{\tintu}{\ifmmode{\rm erg~cm^{-2}~s^{-1}sr^{-1}}\else 
  erg~cm$^{-2}$~s$^{-1}$~sr$^{-1}$\fi}
\newcommand{\fluxu}{\ifmmode{\rm erg~cm^{-2}~s^{-1}}\else 
  erg~cm$^{-2}$~s$^{-1}$\fi}
\newcommand{\velu}{$\,$km$\,$s$^{-1}$}

\newcommand{\wave}{\ifmmode{\lambda} \else$\lambda$\fi}

\newcommand\lta { \mathrel {\hbox to 0pt {\lower 3.7pt \hbox{$\sim$}
      \hss} \raise 1.7pt \hbox{$<$}}}
\newcommand\gta { \mathrel {\hbox to 0pt {\lower 3.7pt \hbox{$\sim$}
      \hss} \raise 1.7pt \hbox{$>$}}}

\newcommand{\lev} [5] {$ {\rm {#1} \,{^#2}{#3^{#4}_{#5}}}$}

\newcommand{\three}{{\lev {3p^4 (^1D) 3d} {2} {G} { \phantom{e}} {7/2}}}
\newcommand{\two}{{\lev {3p^43d} {4} {D} { \phantom{e}} {7/2}}}
\newcommand{\one}{{\lev {3p^43d} {4} {D} { \phantom{o}} {5/2}}}
\newcommand{\zero}{{\lev {3p^5} {2} {P} { {o}} {3/2}}}
\newcommand{\zeroo}{{\lev {3p^5} {2} {P} { {o}} {1/2}}}

\newcommand{\philemail}{judge@ucar.edu}

\newcommand{\figga}{
\begin{figure}[h] 
\epsscale{1.2}
\plotone{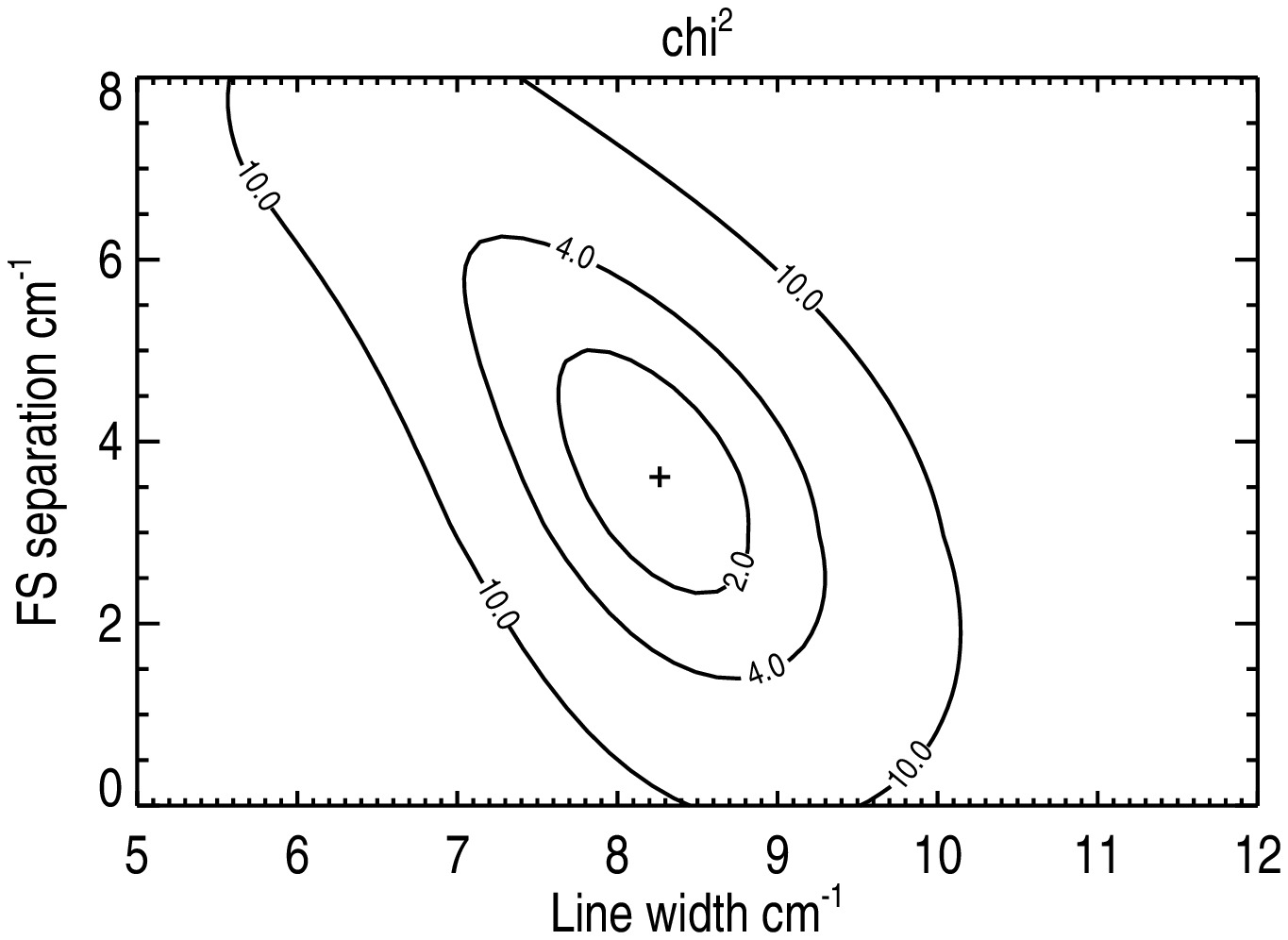}  
\caption{\label{fig:ga} Normalized 
$\chi^{2}$ values are shown as a function of 
line width and FS splitting for a model where the two lines have the same 
line width and an intensity ratio of 2.09:1.  The smallest value is set to
1.   
}
\end{figure} 
}

\newcommand{\figtd}{
\begin{figure}[h] 
\epsscale{1.3}
\plotone{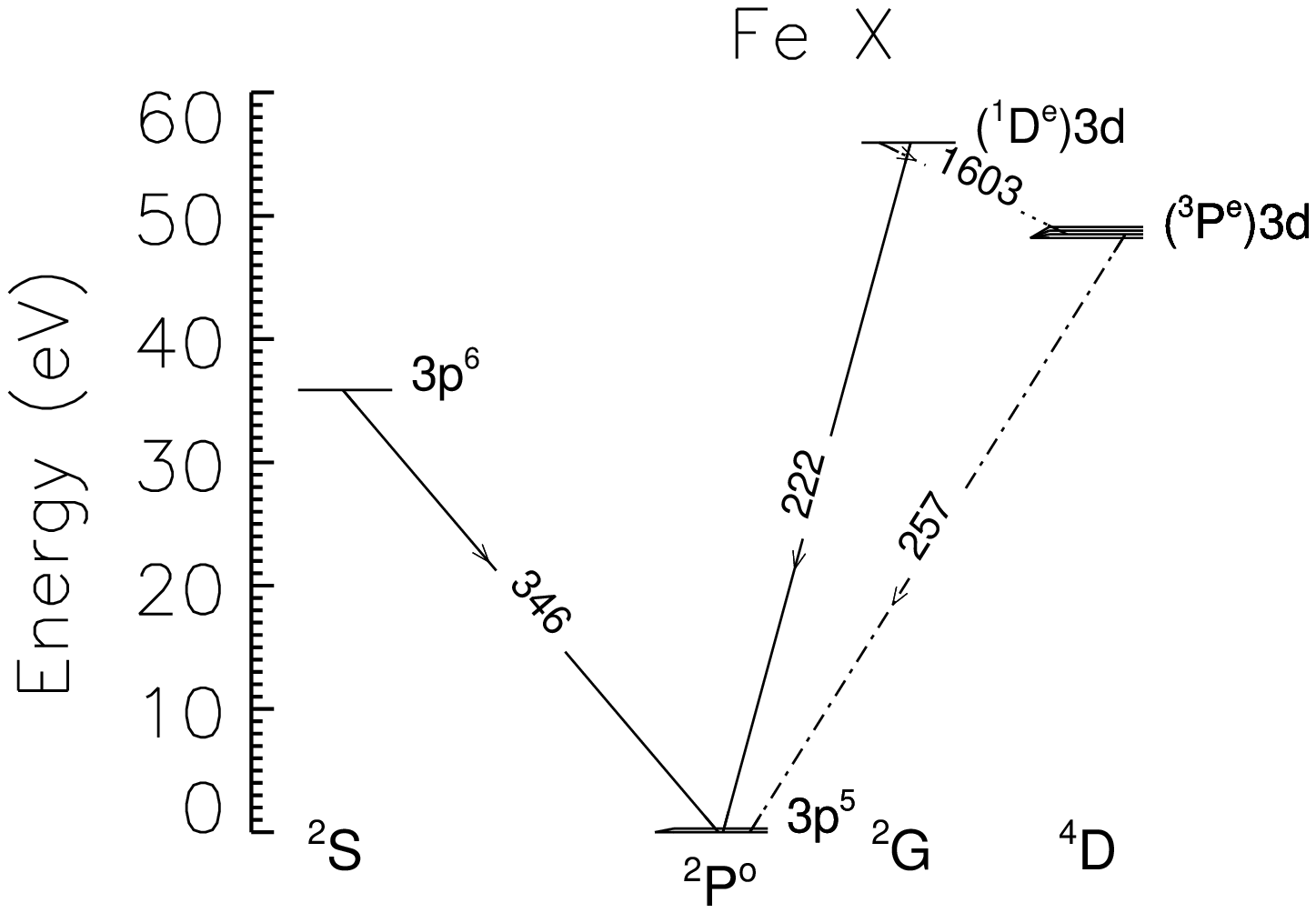}  
\caption{\label{fig:td} A partial term diagram of \ion{Fe}{10} 
showing the levels and (main) transitions of interest.
We make  measurements of differential
wavelengths in the two 1603 \AA{} transitions.
From these data we determine the relative energy
of the lower (${\rm ^4D}$) levels with an accuracy far greater than in the EUV transitions.
}
\end{figure} 
}

\newcommand{\figim}{
\begin{figure*}[h] 
\epsscale{1.}
\plotone{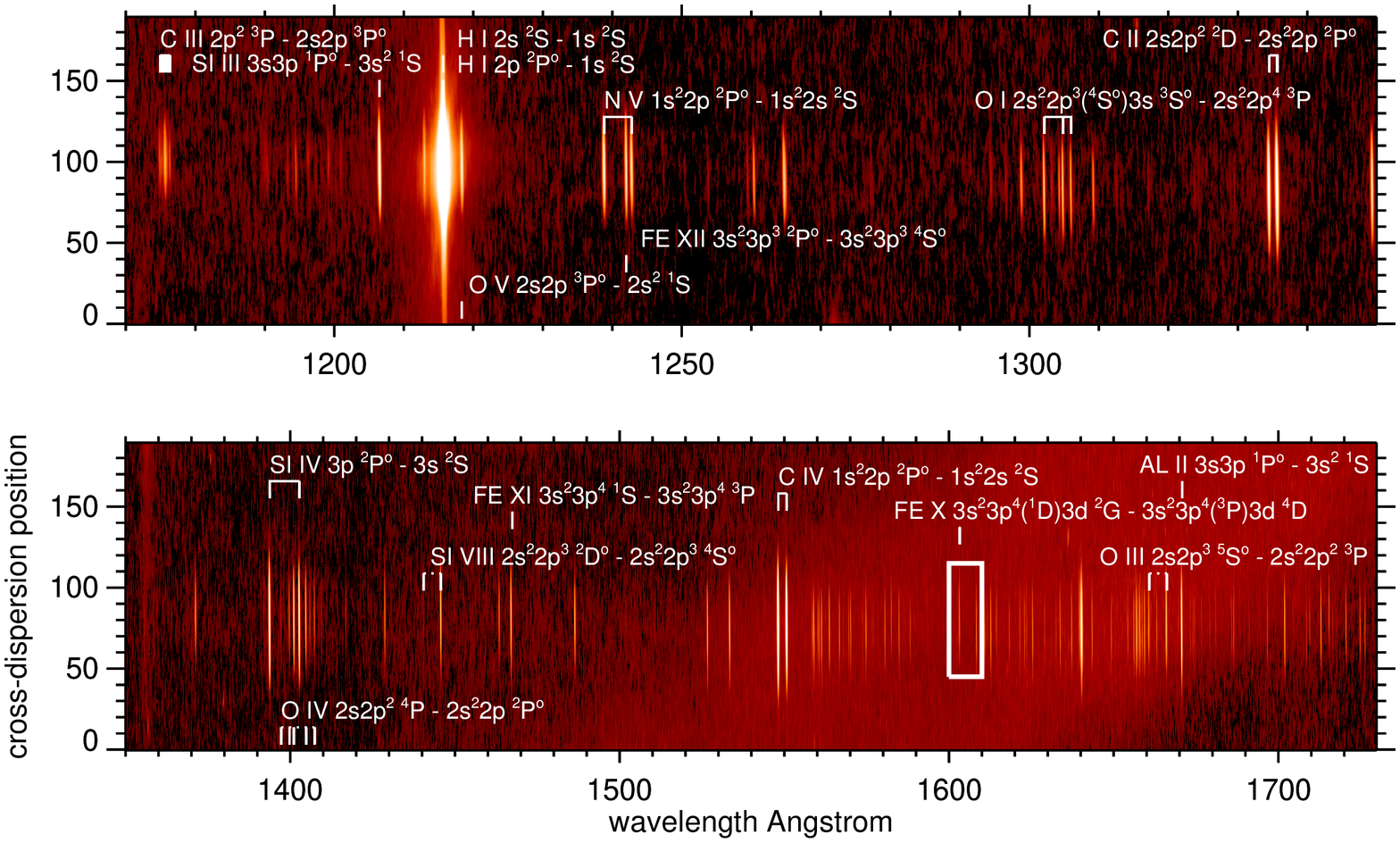}  
\caption{\label{fig:im} Raw S082B image with wavelength scale calibrated as described 
in the text.  The image shown is an off-limb exposure, catalog number  3B158\_007, a
19 minute 59 second exposure obtained on January 9th 1974 beginning at 18:41 UT.  
The ``intensity'' scale shown on the right is photographic density.  The images of the
spectrograph slit are seen along the abscissa. But because the slit was not in a solar image 
plane, there is no solar information in the slit direction.  }
\end{figure*} 
}

\newcommand{\figimc}{
\begin{figure}[ht] 
\epsscale{1.3}
\plotone{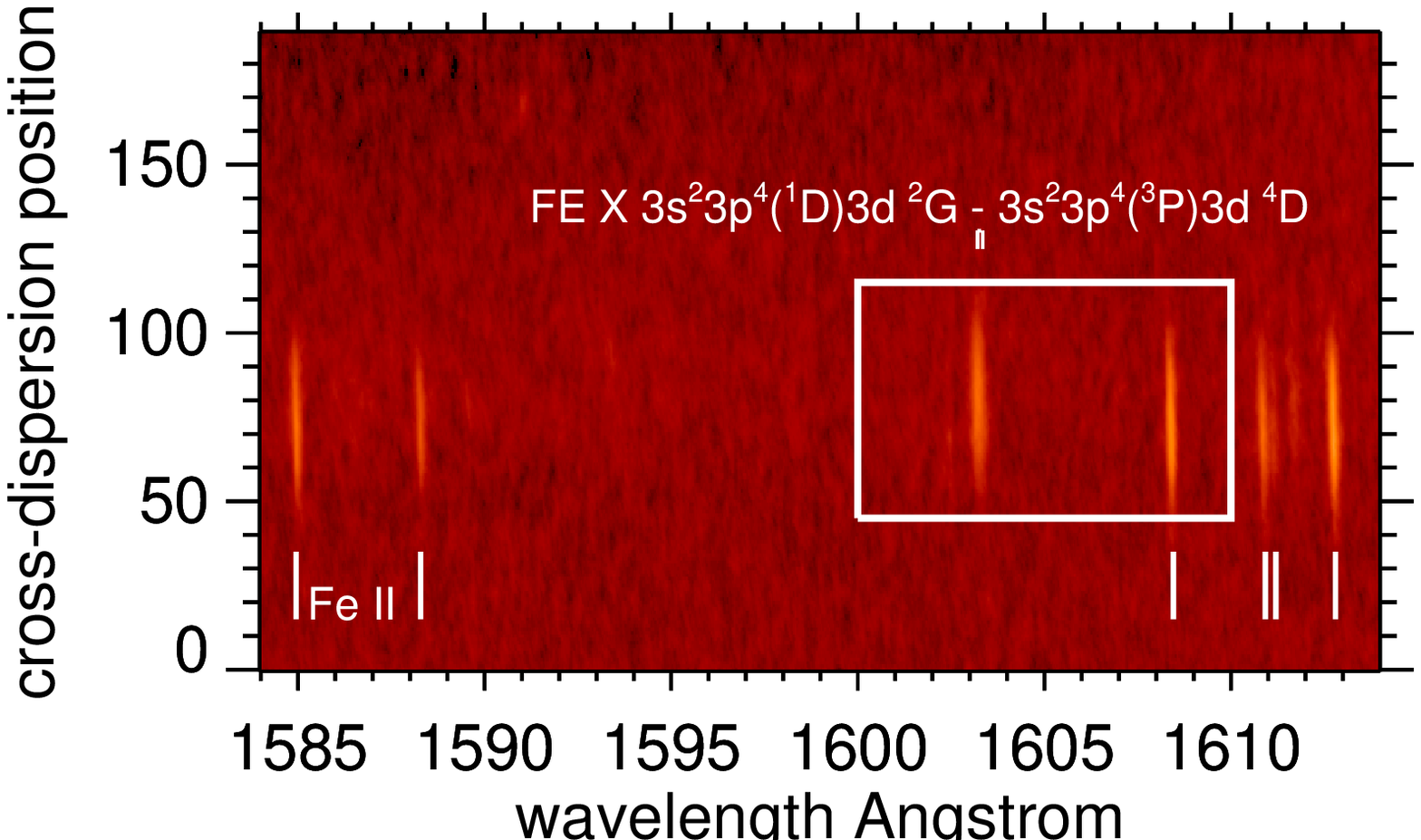}  
\caption{\label{fig:imc} A close-up of the region containing a record 
of the \ion{Fe}{10} lines of interest.  Several lines of \ion{Fe}{2} are seen
in this figure.   }
\end{figure} 
}

\newcommand{\figfit}{
\begin{figure}[h] 
\epsscale{1.1}
\plotone{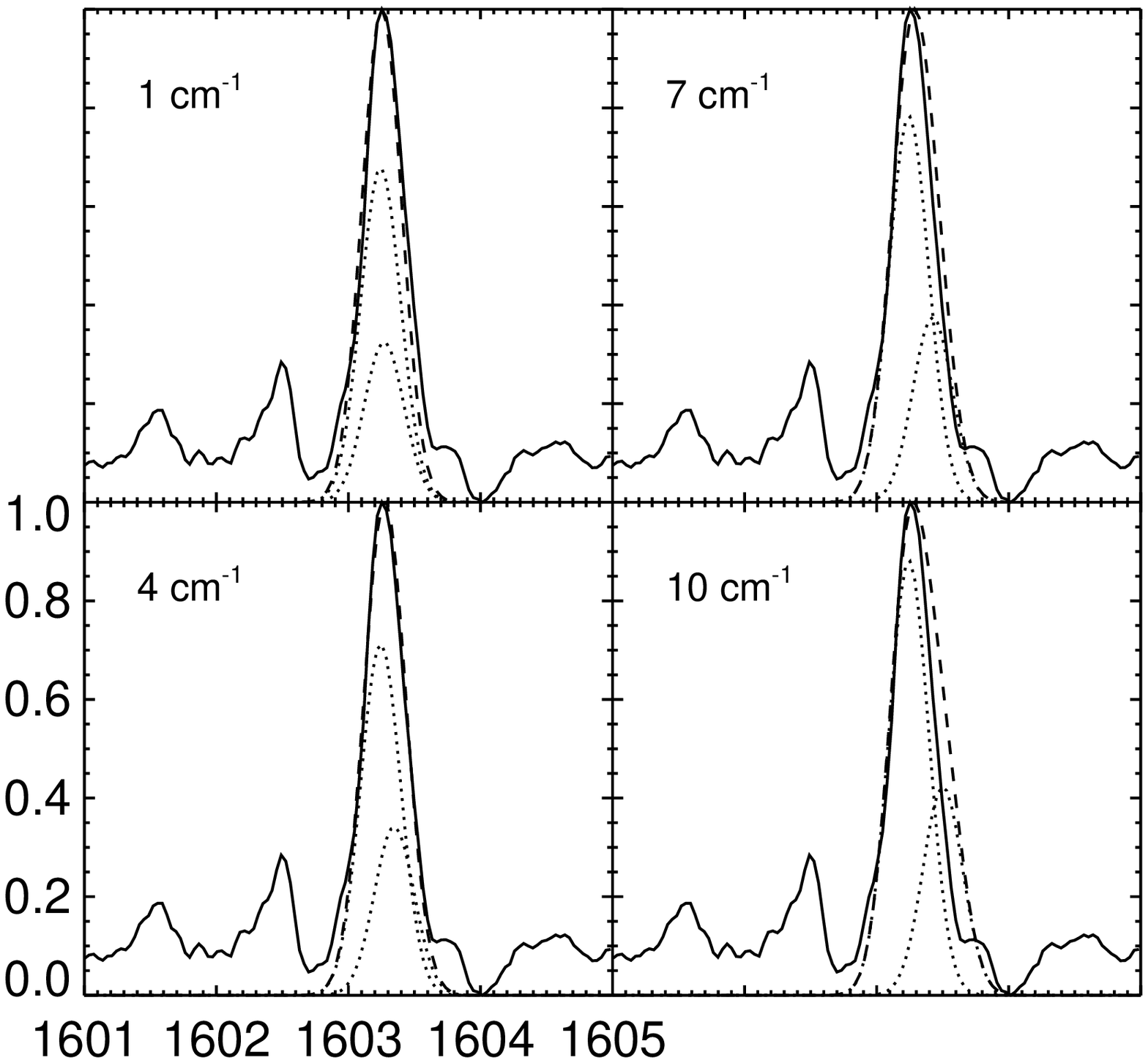}  
\caption{\label{fig:fit} Observed profiles and fits are shown for a
  specified line width of 0.23 \AA, of the 1603.2 \AA{} region of the
  S082B data obtained above the solar limb.  The solid line shows the
  S082B data, the dotted lines the two components of the \ion{Fe}{10}
  transitions in their optically thin ratio, the dot-dashed lines the
  total modeled emission for the given FS splitting.  The panel
  labeled ``4 \icm'' shows an acceptable by-eye  fit, yielding a FS splitting
  close to 4 \icm.  The other panels show the fits, with residuals at
  least 10 times higher, for splitting values of 1, 7, and 10 
  \icm{}. For values of FS splitting near and below the line width
  ($\equiv 8$ \icm) there is some redundancy between FS splitting and
  line width.  }
\end{figure} 
}

\newcommand\tableone{
\begin{table}
\protect\label{tab:fex}
\begin{center}
 \caption{Atomic data for Fe X}
 \begin{tabular}{lrr}     
 \hline
 \hline
Level &    Designation &  Energy  cm$^{-1}$\\    
 \hline
   0 &     $      3s^2 3p^5 {\rm \  ^2P^o_{3/2}} $ &          0.0  \\
   1 &     $      3s^2 3p^5 {\rm \  ^2P^o_{ 1/2}} $ &      15683.1\\
   2 &     $       3s 3p^6 {\rm \  ^2S_{1/2}} $ &     289236\\
   3 &     $ 3s^23p^4 ({\rm ^3P})  3d {\rm \  ^4D_{7/2}} $ &     388710$^\dag$\\
   4 &     $ 3s^23p^4 ({\rm ^3P})  3d {\rm \  ^4D_{5/2}} $ &     388713.5\\
   5 &     $ 3s^23p^4 ({\rm ^3P})  3d {\rm \  ^4D_{3/2}} $ &     390019\\
   6 &     $ 3s^23p^4 ({\rm ^3P})  3d {\rm \  ^4D_{1/2}} $ &     391554\\
  14 &  $   3s^23p^4 ({\rm ^1D})  3d {\rm \ ^2G_{ 7/2}}$  & 451083 \\
\hline
\end{tabular}
\vskip 12pt
Transitions
\vskip 12pt
 \begin{tabular}{lllllll}     
 \hline
 \hline
Type&  $\lambda$ &  up & lo&        A&          gf&          br\\
    &  \AA{} & & & sec$^{-1}$ \\  
\hline
   IC &       257.259&   4 &  0 & $4.4(6)$ & $2.6(-4)$ & 1\\
   M2  &      257.261$^\dag$&   3 &  0 & 47 & $3.726(-9)$ & 1\\
   MIT &      257.261$^\dag$&   3 &  0 &  $3.1(-4){B^2}^\ddag $& \ldots & 1\\
   IC &       256.398&   5 &  0 & $7.7(6)$ & $3.0(-4)$ & 0.98\\
   IC &       255.393&   6 &  0 & $5.9(6)$ & $1.2(-4)$ & 0.39\\
   M2 &       268.075&   4 &  1 & $1.0(1)$ & $6.7(-10)$ & $2.3(-6)$\\ 
   IC &       267.140&   5 &  1 & $1.3(5)$ & $5.7(-6)$ & 0.017 \\
   IC &       266.049&   6 &  1 & $9.4(6)$ & $2.0(-4)$ & 0.61\\
   M2 &       1603.348&  14 &  4 & 9.2  &2.8(-8)  & 0.14\\
   M2 &       1603.260$\dag$&  14 &  3 & 19.2  &5.9(-8)  &0.29 \\
\hline
\end{tabular}
\end{center}
The notation used for the transitions is $X.Y(Z) \equiv X.Y\times
10^Z$, the quantity ``br'' is the radiative branching ratio.  Data are
from the {\tt CHIANTI} database \citep{CHIANTI}, except for the
following: $^\dag$Energy level set to 3.5 \icm{} from the $J=5/2$
level.  $^\ddag$Computed by summing all $M$ substates from equation
(4) of \citet{Li+others2015}, using a splitting of 3.5 \icm, with $B$
measured in G.  ``IC'' is a spin-forbidden intersystem electric dipole
transition, ``M2'' a magnetic dipole transition, ``MIT'' a
magnetically-induced transition.
\end{table}}

\shortauthors{Philip Judge, Roger Hutton, Wenxian Li, Tomas Brage}
\shorttitle{Fe X fine structure}

\slugcomment{}

\begin{document}

%###############################################################################
%
%     OPENING
%
%###############################################################################

\title{On the fine structure splitting of the \one{}  and \two{} levels of Fe~X}
\author{  
Philip G. Judge }
\affil{High Altitude Observatory,\\
       National Center for Atmospheric Research\thanks{The National %
       Center for Atmospheric Research is sponsored by the %
       National Science Foundation},\\
       P.O.~Box 3000, Boulder CO~80307-3000, USA; \philemail}

\author{Roger Hutton and Wenxian Li}
\affiliation{The Key Laboratory of Applied Ion Beam Physics, Ministry of Education, China; rhutton@fudan.edu.cn}
\affiliation{Shanghai EBIT laboratory, Institute of Modern physics, Fudan University, Shanghai, China}

\and
\author{Tomas Brage}
\affiliation{Division of Mathematical Physics, Department of Physics, Lund University, Sweden; Tomas.Brage@fysik.lu.se}

%###############################################################################
%
%     ABSTRACT
%
%###############################################################################

\begin{abstract}
We study UV spectra obtained with the SO82-B slit spectrograph 
on board {\em SKYLAB} to estimate the fine structure splitting 
of the Cl-like \one{} and \two{} levels of \ion{Fe}{10}.   The splitting is
of interest because the Zeeman effect mixes these levels, producing a 
``magnetically induced transition'' (MIT) from \two{} to \zero{} for modest magnetic 
field strengths characteristic of the active solar corona.   We estimate 
the splitting using the Ritz combination formula applied to two lines in
the UV region of the spectrum close to 1603.2 \AA, which decay
from the level \three{} to these two lower levels.  The MIT and 
accompanying spin-forbidden transition lie 
near 257 \AA.  
By careful inspection of a deep exposure obtained with the S082B instrument 
we derive a splitting of $\lta 7 \pm 3$ \icm{}.  The upper limit arises
because of a degeneracy between the effects of non-thermal line broadening 
and fine-structure splitting for small values of the latter parameter. Although
the data were recorded on photographic film, we solved for optimal values of 
line width and splitting of $8.3\pm0.9$ and $3.6\pm 2.7$ \icm{}.
 \end{abstract}


\section{Introduction}

\citet{Grumer+others2014} and \citet{Li+others2015,Li+others2016} 
have presented a 
novel method by which mixing of atomic states by the Zeeman effect can
in principle be used to determine magnetic field strengths in the solar 
corona.   The method is readily illustrated by non-degenerate 
first order quantum perturbation
theory.  Given a ``pure'' set of atomic states $|\alpha_i J_i\rangle$
with angular momentum quantum numbers $J_i$ and other quantum numbers $\alpha_i$, the Zeeman effect will mix the states to first order 
by addition of the magnetic 
Hamiltonian $H_M = \mu_i  B$ where $B$ is the field induction and $\mu_i$
the magnetic moment of state $i$, such that 
\be{expansion}
|\alpha_j J_j\rangle \approx \sum_i d_i |\alpha_i J_i\rangle 
\ee
where the values of mixing coefficients $d_i, i \ne j$ are proportional
to $H_M$ and inversely proportional to the unperturbed fine structure (FS) splitting 
$E_j-E_i$  \citep[e.g.][eq. 3]{Li+others2015}.  

The spectrum of \ion{Fe}{10} is bright in the Sun's corona, the 
well known ``coronal red line'' at 6376.29 \AA{} (\zeroo{} to 
\zero) is one of the strongest forbidden 
transitions visible during eclipse or with coronagraphs 
\citep{Kiepenheuer1953,Billings1966}. 
Some particular \ion{Fe}{10} transitions in the EUV, which can be seen
clearly against the dim EUV solar disk, have been shown to
be of special interest for the challenging problem of 
measuring the coronal magnetic field \citep{Li+others2015}.  This is 
because a near level-crossing occurs close to Fe$^{9+}$
in the Cl-like iso-electronic
sequence, for the two levels \one{} and \two{}.  Then 
$E_j-E_i$ becomes small enough that one mixing coefficient 
$d_i$ become large enough to produce ``magnetically 
induced transitions'' (MITs) that are otherwise forbidden.  
In \ion{Fe}{10} this occurs  
between the unperturbed levels \two{} and \zero{}.  The
radiative decays of the two levels \one{} and \two{} thereby become
sensitive to the magnetic field in the emitting plasma, which can therefore
be measured simply through
the ratios of the two line intensities.

For this method to be of practical use, the energy difference
$E_j-E_i$ must be known with sufficient accuracy
\citep{Li+others2016}.  The latter authors estimated the value of
$E_j-E_i$ for levels \one{} and \two{} using a known magnetic field and
the observed intensity ratio.  At present, the splitting is believed to
be on the order of 3.5 \icm{} from the Shanghai EBIT
measurement, by measurement of the EUV spectrum with a known magnetic field
\citep{Li+others2016}.  But given the high sensitivity of the 
derived magnetic field to the zero-field splitting, it is important to 
try to constrain the energy splitting through other techniques. 

The purpose of the present paper is simply to determine $E_j-E_i$
independently using the Ritz combination principle applied to some
forbidden transitions observed during the {\em SKYLAB} era using the
S082B slit spectrograph.  Figure~\pref{fig:td} shows a partial term
diagram showing the levels of interest to the present work.  This
diagram is based upon atomic data listed in Table~1.

\figtd

\section{Observations and Analysis}

The S082B spectrograph provided data from a $2\arcsec \times 60
\arcsec$ projected slit with a spectral resolution of 0.06 \AA{} \citep{Bartoe+others1977}. 
We searched the logs of the S082B instrument for deep exposures
obtained above the solar limb, in the short wavelength channel (970
to 1970 \AA).  The need for deep, off-limb data is apparent
when we consider the low expected intensities of the lines decaying
from the \three{} level to the \one{} and \two{} levels for two reasons.  First, the
transitions lie close to 1603.2 \AA{}, where the solar continuum seen
on the disk is bright.  Second, the $L$ and $J$ quantum numbers of the upper
level \three{} differs by three and at least 2 from \zero{} and
\zeroo{}. Collisional excitations to \three{} from the ground levels
are therefore infrequent
\citep[e.g.][]{Seaton1962a,Burgess+Tully1992}.

Several parameters are of relevance to our measurement.  
At
1603.2 \AA{}, corresponding to 62,375 \icm{} wave-numbers, the 0.06 \AA{} spectral resolution 
corresponds to
2.3 \icm wave-numbers.  The thermal line-width is, in Doppler units
$\xi= \sqrt{2kT/m} = 17$ \velu, assuming 
that the Fe$^{9+}$ ion temperature is close to the electron
temperature $T_e \approx 10^6$K ($\equiv 86$ eV) under ionization
equilibrium conditions \citep{Jordan1969,Arnaud+Raymond1992}.
The full-width at half-maximum of the
thermal profile is 0.154 \AA{} at 1603.2 \AA, equivalent to 5.97 \icm.  A quadratic
sum of the instrument resolution and thermal width amounts to
0.164 \AA, or 6.4 \icm.  {Inclusion of 16 \velu{} of non-thermal motion 
\citep{Cheng+Doschek+Feldman1979} yields 0.217 \AA, 8.5 \icm.} 

These numbers indicate that the two transitions of interest 
(\three{} to \two{} and \one) 
will be blended, largely because of the line-widths of
\ion{Fe}{10} lines formed in coronal plasma.  

\figimc

We found just one exposure of 
sufficient quality to perform the necessary measurement,
an exposure from January 9th 1974 beginning at 18:41 UT, 
catalog number  3B158\_007, a
19 minute 59 second exposure.  Other promising exposures proved to be 
have significant chromospheric contamination (3B153\_005 for example) or 
were simply under-exposed (3B160\_004).  

Figure~\pref{fig:imc} shows a magnified view of the 1603 \AA{} region, 
the box shows the same region indicated in
Figure~\pref{fig:im} (shown at the end of this paper) 
showing the mid-section of the scanned plate obtained from the NRL data
repository.  Various spectral lines are shown, including
lines used to determine a relative wavelength scale and other
prominent and well-known UV lines.  

We calibrated the spectrograph's wavelength scale using lines 
of \ion{Si}{3} (1206.510 \AA), \ion{C}{2} (1334.535), \ion{Si}{4}
(1393.755 and 1402.770), \ion{C}{4} (1548.202, 1550.774), \ion{O}{3} (1666.153)
 \ion{Al}{2} (1670.787), with lines of \ion{Fe}{2} 
at 1584.949,1588.286,1608.456,1612.802 \AA{} clustered near 1603 \AA.
(Wavelengths are from the compilation of \citealp{Sandlin+others1986}).
We fitted wavelength $\lambda$ to a second order polynomial in position $x$, with 
the result
\be{wavecal}
\lambda =  958.95258 +    a x +  b x^2,
\ee
where $a=0.034635388$ and $b=-1.6092831\times10^{-10}$. 
with residuals ($\pm 1\sigma$) of 0.02 \AA{} across the entire wavelength range.  The residual
corresponds to an uncertainty of 0.7 \icm{} wave-numbers.  
We assumed that the intensity of the 1603 \ion{Fe}{10} blend lies within the linear part of the 
intensity-density curve, because various lines of \ion{Fe}{2} with the same 
photographic densities appearing to be compatible with optically thin 
intensities.  {The intensities were also derived 
from \ion{Fe}{2} line profiles, assumed to be single Gaussians, using 
photographic density $\rho = \alpha + \beta  I + \gamma  I^2$, and solving for
$a,b$ and $c$ from these lines. Then $I$ was solved at neighboring 
wavelengths for the \ion{Fe}{10} transitions.  This procedure 
reduced slightly the lower intensity values at the 
base of the lines, without changing the results significantly.} 

We do not know widths of \ion{Fe}{10} lines that are
unblended, nor do we know the precise wavelength of \ion{Fe}{10} lines
on the wavelength scale above.  Therefore the unresolved FS can only
be derived as {\em an upper limit} by comparing the observed data with
the two lines, estimating for the FS splitting itself (or by
a least-squares optimization outlined below).  In making these
comparisons, we adopted the optically thin ratio of 2.09 for the two
lines from {\tt CHIANTI} \citep{CHIANTI}.  Some typical profiles 
are illustrated in Figure~\pref{fig:fit}.

\figfit

The figure shows observed profiles over-plotted with models with a
given (1/e) width of 0.23 \AA{} (taken from the optimization
calculation below) for four values of the FS splitting.  The
figure suggests 
a FS splitting 
value of $\approx 4$, certainly $\lta 7$ \icm{}. The
variances,  sum of model minus data squared evaluated over the fit to a 1 \AA{}
wide band 
centered at the line, are 2.5, 3 and 9 
times the optimal value for 1, 7 and 10 \icm{} respectively.

{
\tableone
}
A trade-off exists between line-width and FS splitting: the larger the
width, the lower the splitting needed to fit the same data.
Therefore, we can place only an upper limit on the FS splitting from
this analysis of $\lta 4 \pm 3$ \icm.  But we can go a step further.
While the data are photographic and subject to non-linearities in
photographic density vs. intensity, we nevertheless performed a formal
analysis of variances by fitting the two lines and varying the strong
line's wavelength, line widths and the FS splitting.  Both lines were
given the same width and the ratio of the intensities was fixed at
2.09.  The {\tt pikaia} genetic algorithm was used
\citep{Charbonneau1995a}.  Figure~\pref{fig:ga} shows the $\chi^2$
surface as a function of width and FS splitting.  We adopted a
constant value for the observational error across the line profiles
for the $\chi^2$ calculation, because the photographic densities lie
on top of a large pedestal (equivalent to a dark current).  We
multiplied the errors by Gaussian functions centered near the obvious
blends to either side of the core \ion{Fe}{10} emission, to give
higher weight to the fits nearer the cores of the \ion{Fe}{10} lines
of interest.  We use the locus of contour level 2 (2$\times$ the
minimum $\chi^2$) to estimate the error bars in width and FS
splitting.  We find $8.3\pm0.6$ and $3.6\pm 1.8$ \icm{} respectively,
using this criterion.  These error bars are difficult to justify, they
depend on our assumption on the observational errors.  
We made additional 
experiments with different observational error weightings. 
They might reasonably be $1.5\times$ larger, 
so our final conservative estimates are $8.3\pm0.9$ and $3.6\pm 2.7$ 
respectively.  Lastly, we
found the central wavelength of the stronger transition, on the scale
defined using the mix of chromospheric and transition region lines, to
be 1603.25 $\pm 0.02$ \AA.

\figga

The value of $3.6 \pm 2.7$ \icm{} 
is an independent verification of the small estimate of 
the splitting $\approx 3.5$ \icm{} 
obtained entirely independently by \citet{Li+others2016}
from the Shanghai EBIT device.  Our work rejects some of the larger values 
examined by \citet{Li+others2015}. The small value of the splitting found here
also confirms that the line ratios identified by \citet{Li+others2015} {\em can
in principle be used to derive interesting 
values of the coronal magnetic field strength
over active regions.}

{Finally, we remind the reader that diagnosis of magnetic fields with the MIT technique 
faces challenges of blended lines and of dependence of line ratios on
plasma density \citep[see Figure 6 of][] {Li+others2015}.  Work is in preparation discussing
these issues.  }

\acknowledgments
We thank the referee for a useful and thoughtful report, and P. Bryans for his
comments.

\figim


\begin{thebibliography}{}

\bibitem[\protect\astroncite{Arnaud and Raymond}{1992}]{Arnaud+Raymond1992}
Arnaud, M. and Raymond, J.: 1992,
\newblock {\em Astrophys.\ J.\/} {\bf 398}, 394

\bibitem[\protect\astroncite{{Bartoe} {\em et~al.}}{1977}]{Bartoe+others1977}
{Bartoe}, J.-D.~F., {Brueckner}, G.~E., {Purcell}, J.~D., and {Tousey}, R.:
  1977,
\newblock {\em Applied Optics\/} {\bf 16}, 879

\bibitem[\protect\astroncite{{Billings}}{1966}]{Billings1966}
{Billings}, D.~E.: 1966,
\newblock {\em {A guide to the solar corona}\/},
\newblock Academic Press, New York

\bibitem[\protect\astroncite{Burgess and Tully}{1992}]{Burgess+Tully1992}
Burgess, A. and Tully, J.~A.: 1992,
\newblock {\em Astron.\ Astrophys.\/} {\bf 254}, 436

\bibitem[\protect\astroncite{{Charbonneau}}{1995}]{Charbonneau1995a}
{Charbonneau}, P.: 1995,
\newblock {\em Astrophys.\ J.\ Suppl.\ Ser.\/} {\bf 101}, 309

\bibitem[\protect\astroncite{{Cheng} {\em
  et~al.}}{1979}]{Cheng+Doschek+Feldman1979}
{Cheng}, C.-C., {Doschek}, G.~A., and {Feldman}, U.: 1979,
\newblock {\em Astrophys.\ J.\/} {\bf 227}, 1037

\bibitem[\protect\astroncite{{Grumer} {\em et~al.}}{2014}]{Grumer+others2014}
{Grumer}, J., {Brage}, T., {Andersson}, M., {Li}, J., {J{\"o}nsson}, P., {Li},
  W., {Yang}, Y., {Hutton}, R., and {Zou}, Y.: 2014,
\newblock {\em Physica Scripta\/} {\bf 89(11)}, 114002

\bibitem[\protect\astroncite{Jordan}{1969}]{Jordan1969}
Jordan, C.: 1969,
\newblock {\em Mon.\ Not.\ R.\ Astron.\ Soc.\/} {\bf 142}, 501

\bibitem[\protect\astroncite{Kiepenheuer}{1953}]{Kiepenheuer1953}
Kiepenheuer, K.~O.: 1953,
\newblock in G.~P. Kuiper (Ed.), {\em The Sun\/}, Chicago University Press,
  Chicago, p.~322

\bibitem[\protect\astroncite{{Li} {\em et~al.}}{2015}]{Li+others2015}
{Li}, W., {Grumer}, J., {Yang}, Y., {Brage}, T., {Yao}, K., {Chen}, C.,
  {Watanabe}, T., {J{\"o}nsson}, P., {Lundstedt}, H., {Hutton}, R., and {Zou},
  Y.: 2015,
\newblock {\em Astrophys.\ J.\/} {\bf 807}, 69

\bibitem[\protect\astroncite{{Li} {\em et~al.}}{2016}]{Li+others2016}
{Li}, W., {Yang}, Y., {Tu}, B., {Xiao}, J., {Grumer}, J., {Brage}, T.,
  {Watanabe}, T., {Hutton}, R., and {Zou}, Y.: 2016,
\newblock {\em Astrophys.\ J.\/} {\bf 826}, 219

\bibitem[\protect\astroncite{Sandlin {\em et~al.}}{1986}]{Sandlin+others1986}
Sandlin, G.~D., Bartoe, J.-D.~F., Brueckner, G.~E., Tousey, R., and VanHoosier,
  M.~E.: 1986,
\newblock {\em Astrophys.\ J.\ Suppl.\ Ser.\/} {\bf 61}, 801

\bibitem[\protect\astroncite{Seaton}{1962}]{Seaton1962a}
Seaton, M.~J.: 1962,
\newblock in D.~R. Bates (Ed.), {\em Atomic and Molecular Processes\/},
  Academic Press, New York, ~11

\bibitem[\protect\astroncite{{Young} {\em et~al.}}{2016}]{CHIANTI}
{Young}, P.~R., {Dere}, K.~P., {Landi}, E., {Del Zanna}, G., and {Mason},
  H.~E.: 2016,
\newblock {\em Journal of Physics B Atomic Molecular Physics\/} {\bf 49(7)},
  074009

\end{thebibliography}
\end{document}